\documentclass[prl,twocolumn,showpacs,
preprintnumbers,amsmath,amssymb,tightenlines]{revtex4}
\usepackage{graphicx}
\newcommand{\beq}{\begin{equation}}
\newcommand{\eeq}{\end{equation}}
\newcommand{\bea}{\begin{eqnarray}}
\newcommand{\eea}{\end{eqnarray}}
\newcommand{\ba}{\begin{array}}
\newcommand{\ea}{\end{array}}
\newcommand{\bc}{\begin{center}}
\newcommand{\ec}{\end{center}}

\newcommand{\bml}{\begin{mathletters}}
\newcommand{\eml}{\end{mathletters}}
\newcommand{\commentout}[1]{{}}
\newcommand{\bk}{{\bf k}}

\newcommand{\half}{\hbox{$1\over2$}}

\newcommand{\eq}[1]{(\ref{#1})}
\newcommand{\etal} {{\it et al.\/}}
\newcommand{\comment}[1]{{}}

\begin{document}
\title{Collective molecule formation in a
degenerate Fermi gas via a Feshbach
Resonance}
\author{Juha Javanainen}
\author{Marijan Ko\u{s}trun}
\author{Yi Zheng}
\affiliation{Department of Physics, University of
Connecticut, Storrs, Connecticut
06269-3046}
\author{Andrew Carmichael}
\author{Uttam Shrestha}
\author{Patrick J. Meinel}
\affiliation{Department of Physics, University of
Connecticut, Storrs, Connecticut
06269-3046}
\author{Matt Mackie}
\altaffiliation [Also at]{
Department of Physics, University of Turku,
FIN-20014 Turun
yliopisto, Finland.}
\affiliation{
Helsinki Institute of Physics, PL 64, FIN-00014
Helsingin yliopisto,
Finland}
\author{Olavi Dannenberg}
\affiliation{
Helsinki Institute of Physics, PL 64, FIN-00014
Helsingin yliopisto,
Finland}
\author{Kalle-Antti
Suominen}
\altaffiliation [Also at]{
Department of Physics, University of Turku,
FIN-20014 Turun
yliopisto, Finland.}
\affiliation{
Helsinki Institute of Physics, PL 64, FIN-00014
Helsingin yliopisto,
Finland}

\begin{abstract}
We model collisionless collective conversion of a
degenerate Fermi gas into
bosonic molecules via a Feshbach resonance, treating
the bosonic
molecules as a classical field and seeding the pairing
amplitudes with random
phases. A dynamical instability of the Fermi sea
against association into
molecules initiates the
conversion. The model qualitatively reproduces several
experimental observations {[Regal et al.,
Nature {\bf 424}, 47 (2003)]}.
We predict that the initial temperature of the
Fermi gas sets the limit for the efficiency of
atom-molecule conversion.
\end{abstract}
\maketitle


The idea~\cite{TIM98} that an adiabatic sweep across
an
atom-molecule resonance can transform an atomic
condensate into a molecular
condensate has recently been ported to experiments on
degenerate Fermi
gases. By sweeping a magnetic field across a Feshbach
resonance, at least part of the atoms have been
demonstrably converted into
molecules~\cite{REG03,STR03}. Magnetoassociation of
atoms into
molecules via a Feshbach resonance is also the key to
experiments in which formation of a molecular
condensate out of a degenerate
Fermi gas has been observed~\cite{GRE03}.

To date, most experiments on magnetoassociation of
fermionic atoms into molecules
have been done in the collision-dominated regime.
Collisions induce thermal
equilibrium, and statistical mechanics, or indeed
thermodynamics, seems to be the
appropriate theoretical framework~\cite{CAR03} (see
also Cubizolles et
al.~\cite{STR03}). An obvious exception is the
adiabatic-sweep
experiments of
Ref.~\cite{REG03} on fast enough time scales that
particle collisions are not a major
factor~\cite{FOOT1}. These experiments are the domain
of the time-dependent Schr\"odinger equation.
Time-dependent association of a Bose-Einstein
condensate of atoms into a condensate of molecules has
attracted much interest~\cite{TIM98,KOK02a}. Treating
the condensates as
classical fields, as opposed
to quantum fields, guides and simplifies the analysis
of Bose systems. In contrast, even the
zero-temperature Fermi sea of atoms
presumably cannot be represented as a classical field,
a ``macroscopic wave
function''. This is the technical dilemma we set out
to tackle.

Here we develop a collisionless model for
magnetoassociation of a
two-component Fermi gas into bosonic molecules,
treating the boson field
classically. In this setting atom-molecule
conversion builds up from a dynamical instability. We
report on
comparisons with experiments~\cite{REG03}, and
make the testable prediction that
temperature limits the conversion efficiency in an
adiabatic sweep of the
magnetic field across the Feshbach resonance.

Consider a free two-component Fermi gas (spin-up and
spin-down) with
the annihilation operators $c_{\bk\uparrow}$ and
$c_{\bk\downarrow}$ for
states with momentum $\hbar \bk$, and the
corresponding
Bose gas of diatomic
molecules with annihilation operators $b_\bk$. Absent
collisions, the Hamiltonian
reads
\begin{eqnarray}
{H\over\hbar} &=& \sum_\bk\left[\epsilon_\bk
(c^\dagger_{\bk\uparrow}c_{\bk\uparrow}+
c^\dagger_{\bk\downarrow}c_{\bk\downarrow})+(\delta+\half\epsilon_\bk
)b^\dagger_\bk b_\bk
\right]\nonumber\\
&+&\sum_{\bk\bk'}
\left[\kappa_{\bk\bk'}c^\dagger_{\bk\uparrow}c^\dagger_{\bk'\downarrow}
b_{\bk+\bk'}
+
{\rm H.c.}
\right]\,.
\label{HAM}
\end{eqnarray}
Here
$\hbar\epsilon_\bk\equiv\hbar\epsilon_k=\hbar^2k^2/2m$
is the kinetic
energy for an atom with mass $m$, $\hbar\delta$ is the
atom-molecule energy
difference that is adjusted by
varying the magnetic field, and $\kappa_{\bk\bk'}$ are
matrix elements for
combining two atoms into a molecule. For $s$-wave
processes,  $\kappa_{\bk\bk'}$ are functions
of the relative kinetic energy of an atom pair.
The Hamiltonian~\eq{HAM} conserves the invariant
particle number
$N=\sum_\bk(c^\dagger_{\bk\uparrow}c_{\bk\uparrow}+
c^\dagger_{\bk\downarrow}c_{\bk\downarrow}+2
b^\dagger_\bk b_\bk)$. Given the
quantization volume $V$, the invariant density is
$\rho=N/V$.

In the spirit of classical field theory, the main
assumption of our model is
that the boson operators in the Heisenberg equations
of motion are declared to be
complex numbers. To facilitate the numerics, we
furthermore only keep the molecular mode with $\bk=0$.
We also assume that
initially the occupation numbers of the spin-up and
spin-down fermions are the
same, and that the sample is rotationally invariant.
The expectation values of the
relevant operators then depend only on the energy of
the
state $\bk$,
$\hbar\epsilon=\hbar^2k^2/2m$. We scale the $c$-number
molecular amplitude as
$\beta\equiv \sqrt{2/N}\, b_0$, define the fermion
occupation numbers
$P(\epsilon)=\langle
c^\dagger_{\bk\uparrow}c_{\bk\uparrow}\rangle=\langle
c^\dagger_{\bk\downarrow}c_{\bk\downarrow}\rangle$,
the
pairing or ``anomalous''
amplitudes $C(\epsilon) = \langle
c_{-\bk\downarrow}c_{\bk\uparrow} \rangle$, and
find
\bea
i\dot{C}(\epsilon)&=&2\epsilon\,C(\epsilon)
+
   \hbox{$1\over\sqrt2$}{\Omega}\label{BEQEQ}\,
f(\epsilon)\beta[1-2P(\epsilon)]\,,\\
i\dot{P}(\epsilon)&=&
\hbox{$1\over\sqrt2$}\,\Omega\,f(\epsilon)[\beta
C^*(\epsilon)-\beta^*C(\epsilon)]\,,\\
i\dot\beta &=& \delta\beta+ \frac{3\,\Omega }
    {{2\sqrt{2}}\,{{{\epsilon }_F}}^{{3}/{2}}}
\int
d\epsilon\,\sqrt{\epsilon}\,f(\epsilon)C(\epsilon)\,.\label{INTEGRAL}
\label{FINEQ}
\eea
Here $\hbar\epsilon_F=\hbar^2(3\pi^2\rho)^{2/3}/2m$ is
the usual Fermi
energy. The energy-dependent atom-molecule coupling
is $\kappa(\epsilon)=\kappa(0) f(\epsilon)$ with
$f(0)=1$, and
the Rabi-like frequency is
$\Omega=\sqrt{N}\kappa(0)$. As per Javanainen and
Mackie~\cite{TIM98},
$\kappa(0)\propto1/\sqrt{V}$, so that
$\Omega\propto\sqrt\rho$. The integral arises from the
continuum limit of the
sum over $\bk$, and $\sqrt{\epsilon}$ is a phase space
factor responsible for the Wigner threshold law for
the dissociation rate
of molecules into atoms.

The problem with Eqs.~(\ref{BEQEQ}-\ref{FINEQ}) is
that
$\beta(t)=C(\epsilon,t)\equiv0$
is always a solution. However, this
solution may be unstable. To illustrate, we carry out
the linear stability
analysis of Eqs.~(\ref{BEQEQ}-\ref{FINEQ}) around the
trivial
solution for given
occupation numbers $P(\epsilon)$. In the usual
single-pole
approximation that becomes increasingly accurate in
the limit
$\Omega\rightarrow0$, and for $\delta>0$, the Fourier
transform of the small deviation from
$\beta=C(\epsilon)=0$ has the pole
\begin{equation}
\omega_0=\delta-i\,{3\pi\Omega^2\over8\epsilon_F^2}\sqrt{\delta\over2}\,
\left|f\left({\delta\over2}\right)\right|^2\,
\left[
1-2P\left({\delta\over2}\right)
\right]\,.
\end{equation}
If the fermion occupation number satisfies
$P(\epsilon)>\half$ for
some energy $\hbar\epsilon$, for a suitable detuning
$\delta$ the
evolution frequency has a positive imaginary part.
  The
implication is  that a
small perturbation from the stationary state
$\beta=C(\epsilon)=0$ grows exponentially.

If dissociation of an isolated molecule
into two atoms is energetically possible, it will
invariably happen because the
state space for two atoms is much larger than for
a molecule. On the other
hand, a filled Fermi sea of atoms may block
dissociation. The state space of allowed molecular
states is then
the one that is larger, and the atoms are prone to
{\em spontaneous
magnetoassociation\/} into molecules. This is the
nature of the
instability.

The Fermi sea is thermodynamically unstable against
formation of Cooper pairs~\cite{COO56}, and resonance
superfluids~\cite{TIM01} inherit an analog of this
trait of BCS superconductors. Nonetheless,
suggestive as the similarity may be, the BCS
instability is different from the present one.
Thermodynamical instability  occurs because pairing
lowers the energy, and so coupling to a reservoir with
a low enough temperature leads to pairing. The
hallmark of dynamical instability is that a small
perturbation grows exponentially in time, environment
or no environment.

Quantum fluctuations could trigger spontaneous
magnetoassociation, but they are
absent in our model. While much is known about
modeling of quantum
fluctuations classically in boson system~\cite{GAR00},
no corresponding general
methods exist for fermions. We resort to the following
heuristic device. Instead of
starting out a calculation with zero anomalous
amplitudes,  we initially seed them with random
numbers having zero average; specifically, nonzero
numbers with a random phase,
$\langle c_{-\bk\downarrow}c_{\bk\uparrow}
\rangle=e^{\phi_\bk}
\langle c^\dagger_{\bk\uparrow}c_{\bk\uparrow}
\rangle^{1/2}
\langle c^\dagger_{\bk\downarrow}c_{\bk\downarrow}
\rangle^{1/2}$. We then integrate
Eqs.~(\ref{BEQEQ})-(\ref{FINEQ}), whereupon the
initial instability and the subsequent dynamics run
their courses. We do the calculations for many choices
of the random phases
$\phi_\bk$, and average the
results. This procedure correctly reproduces the
initial
evolution $\propto t^2$ of the expectation value of
the number of molecules in
the full quantum model. The approximation lies in
using it at all times. As a technical detail, in our
numerical calculations we
discretize
$P(\epsilon)$ and
$C(\epsilon)$ at equidistant points of $\epsilon$
separated by
$\Delta\epsilon$,
and resort to the analogous process.

\begin{figure}
\hspace{-0.5cm}\includegraphics[width=8.0cm]{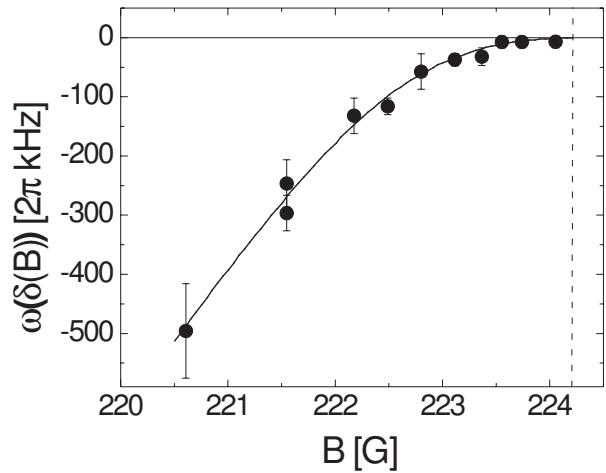}
\caption{Energy of the bound state of the dressed
molecule $\hbar\omega$  as a
function of the magnetic field $B$ from the
experiments
of Ref.~\protect\cite{REG03}
(filled circles), and from our
calculations using the best-fit
parameters $M$, $\Xi$, and $\Delta\mu$ (solid line).}
\label{PARMFIT}
\end{figure}

Next we estimate the parameters of the model. First,
atom-molecule conversion is
unlikely to be the result of a contact interaction.
Given a nonzero range, there
is a cutoff in momentum/energy for the matrix
element $\kappa_{\bk\bk'}$.
We crudely assume that the coupling between atoms
and molecules
follows the Wigner threshold law up to the point when
it abruptly cuts off at some
energy $\hbar M$, and correspondingly set
$f(\epsilon)=\theta(M-\epsilon)$. Second, we write the
dependence of the detuning
on the magnetic field  as $\delta=\Delta\mu
(B-B_0')/\hbar$, where $B_0'$ is a
tentative position of the Feshbach resonance and
$\Delta\mu$ stands for the
thus far unknown difference of the magnetic moment
between a molecule and two
atoms. Third, we have an atom-molecule coupling
strength
with the
dimension of frequency,
$\Xi$, defined by
$\Omega=\Xi^{1/4}\hbar^{3/4}\sqrt{\rho}/m^{3/4}$.

We ignore the Fermi
statistics by setting $P(\epsilon)\equiv0$ in
Eqs.~(\ref{BEQEQ}-\ref{FINEQ}). This yields a linear
description of the coupling between molecules and atom
pairs. With
$\Delta\mu>0$ and for detunings less than a threshold
value
$\delta_0$, the remaining set of equations has a
stationary solution; the Fourier
transforms $\beta(\omega)$ and $C(\epsilon,\omega)$
have a real pole at a frequency
$\omega(\delta)$ such that
$\omega(\delta_0)=0$ and $\omega(\delta) < 0$ for
$\delta<\delta_0$. What in
the absence of the coupling $\propto\Omega$ were a
``bare'' molecule and a pair
of atoms become a ``dressed'' molecule. We interpret
$\hbar\omega[\delta(B)]$ as the  energy of the
bound state of the dressed molecule for the given
magnetic field, and shift the
value of the resonance field from $B_0'$ to $B_0$ so
that
$\delta_0=0$. The ``renormalized'' $B_0$ should equal
the position of the
Feshbach resonance in the limit of a dilute
gas.

Finally, we fit the unknown parameters $M$,
$\Delta\mu$, and $\Xi$ to best
reproduce the experimental binding energy of the
molecule
reported in Fig.~5 of Ref.~\cite{REG03}. The fit
minimizes the relative error
between the calculated and the measured values. The
parameters are
$M=2\pi\times100\,{\rm kHz}$, $\Delta\mu=0.19\,\mu_B$,
and
$\Xi=2\pi\times 580\,{\rm MHz}$.

By setting $P(\epsilon)=0$ we
have ignored the many-body shift of the Feshbach
resonance. We could include  the shift
by allowing $P(\epsilon)\ne0$. For our fitted
parameters, experimental densities, and untouched
zero-temperature Fermi sea, the shift could
exceed one Gauss toward the direction of
high $B$. The preparation of the sample in the
experiments~\cite{REG03} quoted in our
Fig.~\ref{PARMFIT} alters $P(\epsilon)$ and
modifies the shift in a manner that is difficult to
account for self-consistently. Nonetheless, if we
insert a full one-Gauss shift by hand and fit again,
the most relevant parameter
$\Omega$ increases by just 60 \%. Including or
excluding the many-body shift of the Feshbach
resonance should not make a qualitative
difference in the values of the fitted parameters.

\begin{figure}
\hspace{-0.3cm}\includegraphics[width=8.5cm]{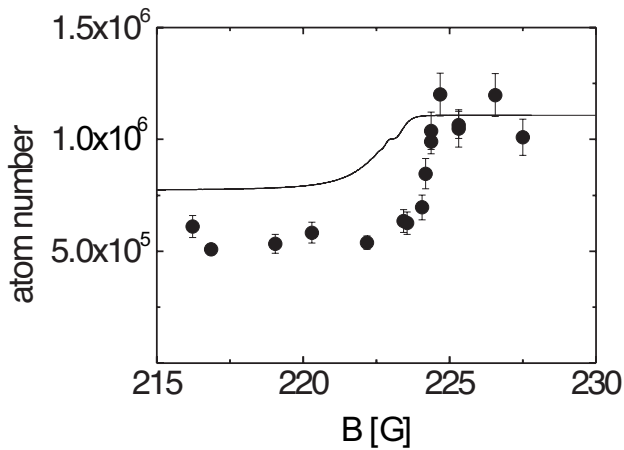}
\caption{Experimental~\protect\cite{REG03} (filled
circles) and
simulated (solid line) numbers of atoms remaining when
the magnetic field is
swept across the Feshbach resonance toward lower
values. The simulations
use the known experimental parameters and parameter
values fitted as in Fig.~\protect\ref{PARMFIT}.}
\label{SWEEP}
\end{figure}

Armed with the parameter values, we next simulate a
sweep of the
magnetic field across the Feshbach resonance as in
Fig.~1(a) of
Ref.~\cite{REG03} by integrating
Eqs.~(\ref{BEQEQ}-\ref{FINEQ}). From the experimental
maximum density
$\rho=2.1\times10^{13}\,{\rm cm}^{-3}$ we have
$\epsilon_F = 9.2\times 2\pi\,{\rm kHz}$. The initial
atomic occupation numbers
$P(\epsilon)$ are set according to the experimental
temperature,
$k_BT/\hbar\epsilon_F = 0.21$. The discretization step
for the atomic
occupation numbers and anomalous amplitudes is
$\Delta\epsilon=\epsilon_F/100$. We run the magnetic
field sweep 64 times for
different random phases of the anomalous amplitudes,
and average the
results. Figure~\ref{SWEEP} shows both the measured
atom numbers (filled circles) and our
calculations (solid line) as a function of the magnetic
field when it is swept at the rate
$(40\, \mu {\rm s/G})^{-1}$.

Experimentally, half of the atoms are converted
into
molecules at $(40\,
\mu {\rm s/G})^{-1}$, while our calculations  give
a 30 \% conversion. There are a number of reasons
why a full quantitative agreement cannot be expected,
e.g., keeping only one
molecular mode, using the maximum density of the atoms
instead of making an
allowance for the density distribution in the trap,
simplistic modeling of the
energy dependence of the atom-molecule coupling
$f(\epsilon)$, and ignoring many-body shifts while
fitting the parameters. Nonetheless, our model seems to
identify the correct physics parameters,
first and foremost the Rabi-like frequency $\Omega$,
and gives a
reasonable estimate for their values.

In the experiments~\cite{REG03} the magnetic field was
also swept back and
forth, whereupon the molecules all dissociate back
into atoms.
With the random initial phases of the anomalous
amplitudes it is not
obvious that our simulations should reproduce this
feat, but they do.

\begin{figure}
\centering
\includegraphics[width=8.0cm]{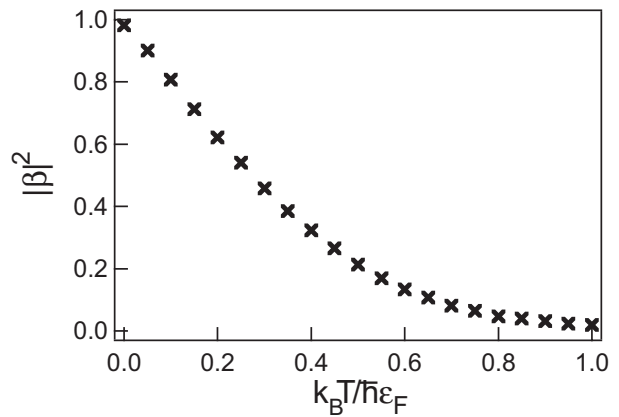}
\caption{Fraction of atoms converted into molecules in
a sweeps such as in
Fig.~\protect\ref{SWEEP}, except that here the sweep
rate is slower,
$(400\,\mu{\rm s/G})^{-1}$, and the results are
plotted as a function of the
varying initial temperature of the atomic gas.}
\label{TLIMIT}
\end{figure}

The puzzling feature of the experiments~\cite{REG03}
is that, no matter how slow the
sweep rate, the conversion efficiency is limited to
about 50\%. To
investigate, we carry out magnetic field sweeps as in
Fig.~\ref{SWEEP}, except that the sweep rate is such
that the detuning as a function of time behaves as
$\delta = -\xi \Omega^2 t$
with $\xi=0.05$, corresponding to $\dot{B}=(400\,\mu
{\rm s/G})^{-1}$. If
$\Omega^{-1}$ determines the time scale,
for $\xi=0.05$ one would expect adiabaticity. We vary
the temperature, and plot
the final conversion efficiency as a function of the
temperature. The results
are shown in Fig.~\ref{TLIMIT}. At $T=0$, 98\% of the
atoms are converted into
molecules. However, by the time the temperature has
reached the Fermi
temperature,
$k_BT=\hbar\epsilon_F$, the conversion efficiency has
dropped to 2\%. At the
typical experimental temperatures with
$k_BT/\hbar\epsilon_F\sim0.2-0.3$, the
efficiency in fact hovers in the neighborhood of 50\%.

For an increasing temperature, a decreasing number of
the initial atomic
states have a thermal occupation number of at least
\half. The instability
responsible for initiating atom-molecule conversion
then acts on fewer and fewer atomic modes. We presume
this is the reason for the temperature
dependence.

The weakest point in our argument may be the
assumption of a single molecular mode, according to
which the
molecules emerge as a
condensate. It would appear
that, without any seed for the molecular condensate or
initial phase from the
atoms, the atoms are converted into a coherent
molecular condensate in a time
that is independent of the size of the sample. This
can hardly be correct.
Inclusion of a multitude of molecular modes would
probably cure this shortcoming.
By construction we would still have a classical field
representing the molecules,
but it could  consist of patches with uncorrelated
phases. A numerical modeling
of this situation is in principle possible, but much
more demanding than our
present calculations. So far we have made little
progress in this direction.

Instead, we present an estimate for the size of the
patches. The
natural velocity parameter in this problem is the
Fermi velocity,
$v_F=\sqrt{2\hbar\epsilon_F/m}$. Suppose the
conversion takes place in a time
$\Delta t$, then atoms in a region of size $\ell\sim
v_F \Delta t$
would plausibly be able to form a patch of molecular
condensate. Let us estimate
the conversion time from Fig.~\ref{SWEEP}, say, as the
difference between the times
when the number of molecules rises from
$1\over4$ to $3\over4$ of its final value in the
dashed-line data, then the size of the patch and the
characteristic distance between the atoms are related
by $\ell \sim 2
\rho^{-1/3}$. Our modeling should be adequate if it is
taken to represent a
patch of about $8^3\sim 10$ atoms. Within each patch
we could still resort to the single
mode-approximation and use the Bose-enhanced coupling
strength $\Omega$ that depends on density but not
directly on the number of atoms. However, the
patchy molecular condensate would not
appear as a condensate in an experiment, but would
display a momentum distribution not narrower than
$\sim\hbar/\ell$.

The number of atoms in a zero-temperature
Fermi sea is the
same as the number of occupied states,
so for
consistency we should
use a step of the order of
$\Delta\epsilon\sim\epsilon_F/10$ in our
modeling. We have used $\Delta\epsilon=\epsilon_F/100$
in
Figs.~\ref{SWEEP} and~\ref{TLIMIT}.  Now,
constants of
the order of unity are beyond our dimensional-analysis
argument,
and the step size estimate should
be taken with a grain
of salt. On the other hand, numerically, the step
size
dependence is logarithmic and
weak. Simulations with just one molecular mode should
therefore be  valid
semiquantitatively.

Starting from a microscopic many-particle theory, we
have modeled collisionless
collective conversion of a degenerate Fermi gas into
bosonic molecules via a
Feshbach resonance. The key techniques are to treat
the molecules
as a classical field, and to seed pairing amplitudes
with random phases. The main
concept we have unearthed is
dynamical instability of a Fermi sea against
magnetoassociation into molecules. The same
instability should occur in
photoassociation. Our model reproduces
qualitatively all
experimental
observations~\cite{REG03} where we have tried a
comparison. Moreover, we have the
testable prediction that temperature sets the limit
for the efficiency of
atom-molecule conversion.

We gratefully acknowledge financial support from NSF
and NASA (UConn), the Magnus
Ehrnrooth Foundation (O.D.), and the Academy of
Finland (M.M. and K.-A.S.,
project 50314).



\begin{references}


\bibitem{TIM98}
    E. Timmermans \etal, cond-mat/9805323;
    J. Javanainen and M. Mackie, Phys. Rev. A {\bf 59},
R3186 (1999);
F. A. van Abeelen and B. J. Verhaar, \prl {\bf 83}
1550 (1999);
    F.\ H.\ Mies, E.\ Tiesinga, and P.\ S.\ Julienne,
\pra
      {\bf 61}, 022721 (2000).

\bibitem{REG03} C.\ A.\ Regal, C.\ Ticknor, J.\ L.\
Bohn, and
D.\ S.\ Jin, Nature {\bf 424}, 47 (2003).

\bibitem{STR03}
    K.\ E.\ Strecker, G.\ B.\ Partridge, and R.\ G.\
Hulet, Phys.\
      Rev.\ Lett.\ {\bf 91}, 080406 (2003);
    J. Cubizolles \etal, \prl {\bf 91}, 240401 (2003).

\bibitem{GRE03}
    S. Jochim \etal, Science {\bf 302}, 2101 (2003);
M. Greiner, C. A. Regal, and D. S. Jin, Nature {\bf
426}, 537
(2003);
    M. W. Zwierlein \etal, \prl {\bf 91}, 250401
(2003).

\bibitem{CAR03} L. D. Carr, G. V. Shlyapnikov, and Y.
Castin, cond-mat/0308306.

\bibitem{FOOT1} Here, of course, we do {\em not\/}
think of the process of
atom-molecule conversion as a collision.

\bibitem{KOK02a}
    S.J.J.M.F. Kokkelmans and M.J. Holland, Phys.\
      Rev.\ Lett.\ {\bf 89}, 180401 (2002);
    M.\ Mackie, K.-A. Suominen, and J.\ Javanainen,
      Phys.\ Rev.\ Lett.\ {\bf 89}, 180403 (2002);
    T.\ K\"ohler, T. Gasenzer, and K.\ Burnett, Phys.\
      Rev.\ A {\bf 67}, 013601 (2003);
    R.A.\ Duine and H.T.C.\ Stoof, J.\ Opt.\ B {\bf 5},
      S212 (2003);
    K.V. Kheruntsyan and P.D. Drummond, \pra {\bf 66}
031602 (2002);
    V.A.\ Yurovsky and A.\ Ben-Reuven, \pra {\bf 67},
050701 (2003).

\bibitem{COO56} L.\ N.\ Cooper,  Phys.\ Rev.\ {\bf
104}, 1189 (1956).

\bibitem{TIM01}
    E.\ Timmermans, K.\ Furuya, P.W.\ Milonni, and
A.K.\
Kerman,
      Phys.\ Lett.\ A {\bf 285}, 228 (2001);
    M.\ Holland, S.J.J.M.F.\ Kokkelmans, M.L.\
Chiofalo,
      and R.\ Walser, Phys.\ Rev.\ Lett.\ {\bf 87},
120406 (2001);
    Y. Ohashi and A. Griffin, Phys. Rev. Lett. {\bf
89},
130402 (2002).

\bibitem{GAR00} C. W. Gardiner and P. Zoller, {\it
Quantum Noise}, 2nd ed.
(Springer, Berlin, 2000).

\end{references}
\end{document}